\documentclass[aps,prd,reprint,nofootinbib,superscriptaddress,floatfix]{revtex4-2}
\usepackage{amsmath,amssymb,bm}
\usepackage{tikz}
\usepackage{pgfplots}
\pgfplotsset{compat=1.16}
\usepackage[colorlinks=true,linkcolor=blue,citecolor=blue,urlcolor=blue]{hyperref}

\newcommand{\keV}{\,\mathrm{keV}}
\newcommand{\MeV}{\,\mathrm{MeV}}
\newcommand{\GeV}{\,\mathrm{GeV}}
\newcommand{\TeV}{\,\mathrm{TeV}}
\newcommand{\kms}{\,\mathrm{km\,s^{-1}}}
\newcommand{\cmsq}{\,\mathrm{cm^{2}}}

\begin{document}

\title{Exothermic dark matter and the 248 keV nuclear recoil in LUX-ZEPLIN}

\author{H. Baer}
\affiliation{Homer L. Dodge Department of Physics and Astronomy, University of Oklahoma, Norman, Oklahoma 73019, USA}

\author{V. Barger}
\affiliation{Department of Physics, University of Wisconsin--Madison, Madison, Wisconsin 53706, USA}

\date{September 10, 2026}

\begin{abstract}
The single 248-keV nuclear recoil reported by LZ is reproduced by dark matter that loses energy when it scatters. The relic is the heavier of two nearly degenerate states, a pseudo-Dirac pair split by $|\delta|\simeq350\keV$, and the collision releases the difference. Many interpretations advanced so far have the dark matter absorb energy, which puts the splitting within a few percent of the xenon kinematic limit; the predicted count then changes from zero to 143 times its central value as the halo escape speed runs from 500 to $600\kms$, while the exothermic count changes by 0.3\%. Only about 1\% of the exothermic events fall below 70 keV, so the null standard search follows. A weak-strength coupling would overproduce the event by $8.9\times10^{5}$, so the mediator is a dark photon, and the excited state survives to the present only for $|\delta|<2m_e$. The relic abundance fixes the dark coupling, and a coupled-channel solution of the conversion between the two states leaves an excited fraction of $2\times10^{-3}$ to $4\times10^{-2}$. One event then requires $\sigma_p\simeq2\times10^{-42}\cmsq$ and a kinetic mixing $\epsilon\simeq10^{-6}$ at $m_{A'}=1\GeV$; beam-dump limits push the mediator above $0.6\GeV$. Upscattering of the ground state through the same operator excludes $|\delta|\lesssim300\keV$, so the published endothermic reading is the small-splitting limit of the same model. The decisive test is argon, whose first form-factor zero lies at 695 keV rather than the 94.6 keV of xenon; argon yields 4 to 6 events per tonne-year near 250 keV, where the endothermic reading is forbidden at any exposure.
\end{abstract}

\maketitle

\section{Introduction}
\label{sec:intro}

LZ's extended-window search, with a $2.84$ tonne-year exposure and nuclear-recoil acceptance from $5.4$ to $269.9\keV$, reports one event consistent with a $248\pm23_{\rm stat}\pm23_{\rm sys}\keV$ xenon recoil, at a maximum local significance of $3.4\sigma$ and a global significance of $2.6\sigma$ after the look-elsewhere correction~\cite{LZhigh}. The analysis is not blind and a single event is not evidence; the interest lies in the recoil energy, which is far above the region where an elastic weakly interacting massive particle would deposit and is instead characteristic of inelastic scattering~\cite{TuckerSmithWeiner}.

The first interpretations advanced were all endothermic~\cite{Yin,Visinelli,FanReece,Freese,SuYangYang,WuZhangZhu}, with a nearly pure Higgsino or a vectorlike electroweak doublet upscattering to a state heavier by $\delta\simeq320$--$360\keV$. In that reading the observed recoil sits at the kinematic endpoint, where the threshold speed $v^\star=\sqrt{2\delta/\mu_A}$ approaches the largest speed available in the laboratory frame. Three conditions must then hold together. The splitting must land within a few percent of the xenon ceiling; the rate must be evaluated in the extreme tail of the velocity distribution, where the standard halo model is least reliable; and, for the Higgsino, $\delta=m_Z^2(s_W^2/M_1+c_W^2/M_2)$ forces the electroweak gauginos to $2.4\times10^{7}\GeV$, which carries a tuned Higgs sector with it. The first two conditions admit a partial defense from selection, since the first event observed in any search sits at its sensitivity frontier, but the sensitivity of the predicted count to halo parameters remains.

Downscattering removes all three. If the relic is the heavier state $\chi_2$ of a pseudo-Dirac pair and the transition observed is $\chi_2A\to\chi_1A$ with the splitting released into the recoil, there is no threshold speed and the recoil energy is set by $|\delta|$ rather than by the halo endpoint. Exothermic dark matter is an established construction~\cite{BatellPospelovRitz,GrahamExo,FinkbeinerLinWeiner}. Two papers that appeared while this work was in preparation also apply it to the LZ event, and the Addendum relates the present analysis to them and to the many other interpretations that have followed the first ones. Section~\ref{sec:kin} gives the kinematics and the xenon form factor, Sec.~\ref{sec:rate} the rate and benchmarks, Sec.~\ref{sec:halo} the sampling of the halo velocity distribution, Secs.~\ref{sec:mediator} and \ref{sec:stability} two constraints that follow from the LZ normalization and from longevity of the relic excited state, Sec.~\ref{sec:solar} the solar bound, Sec.~\ref{sec:tests} the tests, and Sec.~\ref{sec:f2} the relic excited fraction; the Addendum places the reading among the other interpretations of the event.

\section{Downscattering kinematics}
\label{sec:kin}

For a transition of mass splitting $\delta$, with $\delta<0$ denoting energy released, energy and momentum conservation give the threshold speed
\begin{equation}
v_{\min}(E_R)=\frac{1}{\sqrt{2m_AE_R}}\left|\frac{m_AE_R}{\mu_A}+\delta\right|,
\label{eq:vmin}
\end{equation}
with $\mu_A$ the dark-matter--nucleus reduced mass. For $\delta>0$ this has a positive minimum $v^\star=\sqrt{2\delta/\mu_A}$ at $E_R^\star=(\mu_A/m_A)\delta$, and the rate is supported only above $v^\star$. For $\delta<0$ the expression vanishes at $E_R=(\mu_A/m_A)|\delta|$, so every particle in the halo scatters and the recoil spectrum is centered on that value with a width set by the dispersion. At fixed speed the accessible recoils lie between
\begin{equation}
E_R^\pm(v)=\frac{\mu_A^2v^2}{2m_A}\left(1\pm\sqrt{1-\frac{2\delta}{\mu_Av^2}}\right)^2 .
\label{eq:band}
\end{equation}

Two consequences of the target should be stated before the rate. First, the recoil energy is $(\mu_A/m_A)|\delta|$, which tends to $|\delta|$ for $m_\chi\gg m_A$; a $248\keV$ recoil in xenon therefore indicates $|\delta|\simeq275\keV$ at large $m_\chi$ and larger $|\delta|$ for lighter dark matter. Second, the momentum transfer is $q=\sqrt{2m_AE_R}=246\MeV$ at the observed recoil, and the first zero of the Helm form factor~\cite{Helm,LewinSmith} for $^{131}$Xe falls at $q=152\MeV$, i.e.\ $E_R=94.6\keV$. The LZ event therefore lies in the second diffraction lobe, where $|F(q)|^2=1.7\times10^{-4}$ against $0.36$ at $20\keV$. This suppression is common to the endothermic and exothermic readings and is why the cross sections required are far above the elastic limits.

\section{Rate and benchmarks}
\label{sec:rate}

The differential rate per unit detector mass is
\begin{equation}
\frac{dR}{dE_R}=\frac{\rho_0}{m_\chi}\sum_i\frac{x_i}{\overline{m}_A}\,
\frac{\sigma_0^i\,m_{A_i}}{2\mu_{A_i}^2}\,F_i^2(q_i)\,\eta\!\left(v_{\min,i}\right),
\label{eq:rate}
\end{equation}
with $\sigma_0^i=\sigma_p(\mu_{A_i}/\mu_p)^2Z^2$ for a mediator coupled to electric charge, $x_i$ the natural isotope fractions of xenon, and $\eta$ the mean inverse speed of a truncated Maxwellian boosted to the laboratory frame. We adopt the halo parameters used by LZ, $\rho_0=0.3\GeV\,{\rm cm^{-3}}$, $v_0=238\kms$, $v_{\rm esc}=544\kms$, and $v_E=254\kms$, and approximate the efficiency from Fig.~S2 of Ref.~\cite{LZhigh}, rising linearly from $50\%$ at $5.4\keV$ to $96\%$ at $14\keV$, flat to $250\keV$, and falling linearly to $50\%$ at $269.9\keV$. Applied to elastic scattering, the same calculation reproduces the published LZ sensitivity at $1\TeV$~\cite{LZ42} to within a factor of three, so the cross sections quoted below carry that accuracy; the benchmarks are not likelihood fits.

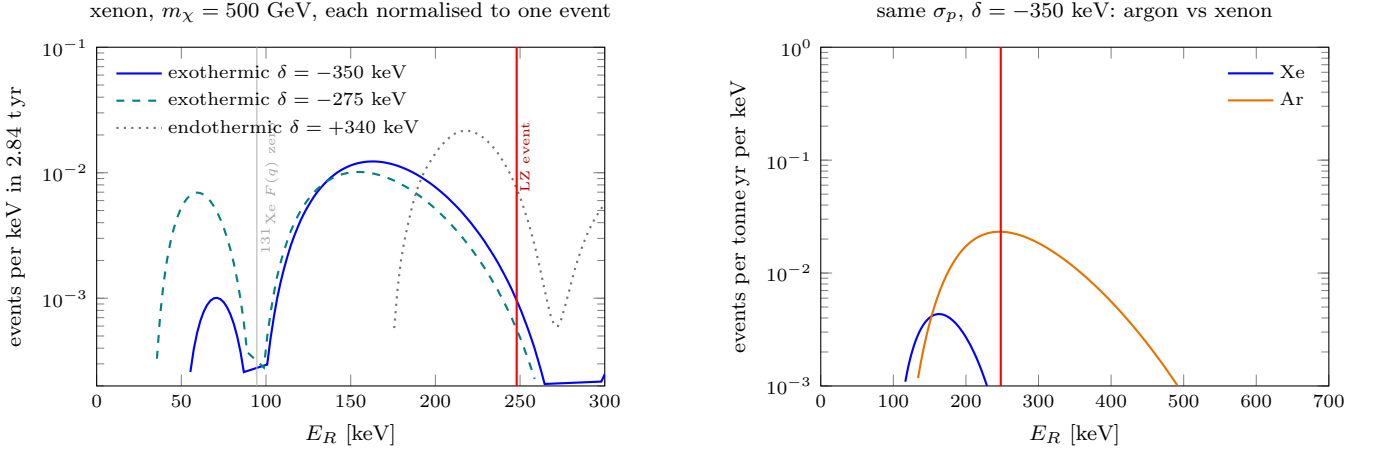
\begin{figure*}[t]
\centering
\begin{tikzpicture}
\begin{axis}[
  width=0.375\textwidth, height=0.25\textwidth, scale only axis,
  xmin=0, xmax=300, ymode=log, ymin=2e-4, ymax=1e-1,
  xlabel={$E_R$ [keV]}, ylabel={events per keV in 2.84 t\,yr},
  title={\footnotesize xenon, $m_\chi=500$ GeV, each normalised to one event},
  label style={font=\footnotesize}, tick label style={font=\scriptsize},
  legend style={font=\scriptsize, draw=none, fill=none, at={(0.02,0.98)}, anchor=north west},
  legend cell align=left, grid=none]
\addplot[blue, thick] coordinates {(55.33,0.00025843) (57.3,0.00039011) (59.28,0.00052804) (61.25,0.00066208) (63.22,0.00078303) (65.19,0.0008831) (67.17,0.0009562) (69.14,0.00099821) (71.11,0.0010071) (73.09,0.000983) (75.06,0.00092798) (77.03,0.00084596) (79.01,0.00074243) (80.98,0.00062408) (82.95,0.00049846) (84.93,0.00037366) (86.9,0.00025792) (100.7,0.00029637) (102.7,0.00048135) (104.7,0.00071744) (106.6,0.0010038) (108.6,0.0013387) (110.6,0.0017194) (112.6,0.0021422) (114.5,0.002603) (116.5,0.0030969) (118.5,0.0036188) (120.4,0.0041631) (122.4,0.004724) (124.4,0.0052959) (126.4,0.0058728) (128.3,0.0064492) (130.3,0.0070196) (132.3,0.0075788) (134.3,0.0081219) (136.2,0.0086445) (138.2,0.0091425) (140.2,0.0096121) (142.1,0.01005) (144.1,0.010453) (146.1,0.01082) (148.1,0.011148) (150,0.011436) (152,0.011683) (154,0.011889) (156,0.012053) (157.9,0.012175) (159.9,0.012256) (161.9,0.012297) (163.9,0.0123) (165.8,0.012264) (167.8,0.012192) (169.8,0.012086) (171.7,0.011947) (173.7,0.011777) (175.7,0.011579) (177.7,0.011355) (179.6,0.011107) (181.6,0.010837) (183.6,0.010547) (185.6,0.01024) (187.5,0.0099178) (189.5,0.0095829) (191.5,0.0092373) (193.4,0.008883) (195.4,0.0085221) (197.4,0.0081565) (199.4,0.007788) (201.3,0.0074183) (203.3,0.0070491) (205.3,0.0066818) (207.3,0.006318) (209.2,0.0059589) (211.2,0.0056058) (213.2,0.0052597) (215.2,0.0049216) (217.1,0.0045925) (219.1,0.004273) (221.1,0.003964) (223,0.003666) (225,0.0033795) (227,0.0031048) (229,0.0028425) (230.9,0.0025926) (232.9,0.0023554) (234.9,0.002131) (236.9,0.0019195) (238.8,0.0017207) (240.8,0.0015346) (242.8,0.0013612) (244.8,0.0012001) (246.7,0.0010513) (248.7,0.00091435) (250.7,0.00078906) (252.6,0.00067508) (254.6,0.00057206) (256.6,0.00047961) (258.6,0.00039734) (260.5,0.00032483) (262.5,0.00026165) (264.5,0.00020737) (298,0.00021667) (300,0.00024865)};
\addlegendentry{exothermic $\delta=-350$ keV}
\addplot[teal, thick, dashed] coordinates {(35.6,0.00032946) (37.57,0.00075382) (39.54,0.0013245) (41.52,0.0020094) (43.49,0.0027696) (45.46,0.0035621) (47.44,0.0043435) (49.41,0.0050727) (51.38,0.0057138) (53.36,0.0062382) (55.33,0.0066254) (57.3,0.0068635) (59.28,0.006949) (61.25,0.0068859) (63.22,0.0066846) (65.19,0.0063606) (67.17,0.0059333) (69.14,0.0054242) (71.11,0.0048562) (73.09,0.0042523) (75.06,0.0036347) (77.03,0.0030241) (79.01,0.002439) (80.98,0.0018958) (82.95,0.001408) (84.93,0.00098646) (86.9,0.00063945) (88.87,0.00037256) (98.74,0.00027992) (100.7,0.00049253) (102.7,0.00077019) (104.7,0.001106) (106.6,0.0014925) (108.6,0.0019221) (110.6,0.0023869) (112.6,0.0028792) (114.5,0.0033915) (116.5,0.0039164) (118.5,0.004447) (120.4,0.0049769) (122.4,0.0055) (124.4,0.006011) (126.4,0.0065048) (128.3,0.0069773) (130.3,0.0074245) (132.3,0.0078432) (134.3,0.0082308) (136.2,0.0085851) (138.2,0.0089045) (140.2,0.0091877) (142.1,0.0094339) (144.1,0.0096429) (146.1,0.0098145) (148.1,0.0099491) (150,0.010047) (152,0.01011) (154,0.010139) (156,0.010134) (157.9,0.010098) (159.9,0.010032) (161.9,0.0099373) (163.9,0.0098166) (165.8,0.0096714) (167.8,0.0095036) (169.8,0.0093151) (171.7,0.0091079) (173.7,0.008884) (175.7,0.0086453) (177.7,0.0083936) (179.6,0.0081308) (181.6,0.0078586) (183.6,0.0075788) (185.6,0.007293) (187.5,0.0070028) (189.5,0.0067096) (191.5,0.0064148) (193.4,0.0061197) (195.4,0.0058256) (197.4,0.0055336) (199.4,0.0052447) (201.3,0.0049599) (203.3,0.00468) (205.3,0.0044059) (207.3,0.0041382) (209.2,0.0038775) (211.2,0.0036245) (213.2,0.0033796) (215.2,0.0031431) (217.1,0.0029154) (219.1,0.0026969) (221.1,0.0024875) (223,0.0022877) (225,0.0020973) (227,0.0019166) (229,0.0017454) (230.9,0.0015838) (232.9,0.0014316) (234.9,0.0012888) (236.9,0.0011552) (238.8,0.0010306) (240.8,0.00091487) (242.8,0.00080772) (244.8,0.00070893) (246.7,0.00061823) (248.7,0.00053535) (250.7,0.00046001) (252.6,0.00039189) (254.6,0.0003307) (256.6,0.00027611) (258.6,0.00022782)};
\addlegendentry{exothermic $\delta=-275$ keV}
\addplot[black!55, thick, dotted] coordinates {(175.7,0.00057851) (177.7,0.0011996) (179.6,0.0019633) (181.6,0.0028188) (183.6,0.0037791) (185.6,0.0049349) (187.5,0.0062145) (189.5,0.0075482) (191.5,0.0088775) (193.4,0.010154) (195.4,0.01134) (197.4,0.012564) (199.4,0.013856) (201.3,0.015137) (203.3,0.016343) (205.3,0.01749) (207.3,0.018578) (209.2,0.019549) (211.2,0.020358) (213.2,0.020975) (215.2,0.02139) (217.1,0.021596) (219.1,0.021588) (221.1,0.021372) (223,0.020957) (225,0.020358) (227,0.0196) (229,0.018721) (230.9,0.017737) (232.9,0.016663) (234.9,0.015517) (236.9,0.014318) (238.8,0.013085) (240.8,0.011839) (242.8,0.010597) (244.8,0.0093781) (246.7,0.0081985) (248.7,0.0070732) (250.7,0.0060151) (252.6,0.0050353) (254.6,0.0041427) (256.6,0.0033444) (258.6,0.0026456) (260.5,0.0020495) (262.5,0.0015575) (264.5,0.0011695) (266.5,0.0008836) (268.4,0.00069669) (270.4,0.00060436) (272.4,0.00060115) (274.3,0.00068068) (276.3,0.00083584) (278.3,0.0010589) (280.3,0.0013418) (282.2,0.001676) (284.2,0.0020531) (286.2,0.0024645) (288.2,0.0029018) (290.1,0.0033566) (292.1,0.0038211) (294.1,0.0042877) (296.1,0.0047493) (298,0.0051995) (300,0.0056322)};
\addlegendentry{endothermic $\delta=+340$ keV}
\draw[gray!60] (axis cs:94.6,2e-4) -- (axis cs:94.6,1e-1);
\node[gray!70, font=\tiny, rotate=90, anchor=north east] at (axis cs:92,3e-2) {$^{131}$Xe $F(q)$ zero};
\draw[red, thick] (axis cs:248,2e-4) -- (axis cs:248,1e-1);
\node[red!70!black, font=\tiny, rotate=90, anchor=north east] at (axis cs:245,3e-2) {LZ event};
\end{axis}
\end{tikzpicture}%
\hfill%
\begin{tikzpicture}
\begin{axis}[
  width=0.375\textwidth, height=0.25\textwidth, scale only axis,
  xmin=0, xmax=700, ymode=log, ymin=1e-3, ymax=1e0,
  xlabel={$E_R$ [keV]}, ylabel={events per tonne\,yr per keV},
  title={\footnotesize same $\sigma_p$, $\delta=-350$ keV: argon vs xenon},
  label style={font=\footnotesize}, tick label style={font=\scriptsize},
  legend style={font=\scriptsize, draw=none, fill=none, at={(0.98,0.98)}, anchor=north east},
  legend cell align=left]
\addplot[blue, thick] coordinates {(116.5,0.0010905) (118.5,0.0012742) (120.4,0.0014659) (122.4,0.0016634) (124.4,0.0018647) (126.4,0.0020679) (128.3,0.0022708) (130.3,0.0024717) (132.3,0.0026686) (134.3,0.0028598) (136.2,0.0030439) (138.2,0.0032192) (140.2,0.0033845) (142.1,0.0035387) (144.1,0.0036808) (146.1,0.0038099) (148.1,0.0039254) (150,0.0040268) (152,0.0041138) (154,0.0041862) (156,0.0042439) (157.9,0.0042869) (159.9,0.0043156) (161.9,0.0043301) (163.9,0.0043308) (165.8,0.0043183) (167.8,0.004293) (169.8,0.0042555) (171.7,0.0042066) (173.7,0.0041469) (175.7,0.0040772) (177.7,0.0039983) (179.6,0.0039108) (181.6,0.0038157) (183.6,0.0037137) (185.6,0.0036056) (187.5,0.0034922) (189.5,0.0033743) (191.5,0.0032526) (193.4,0.0031278) (195.4,0.0030007) (197.4,0.002872) (199.4,0.0027422) (201.3,0.0026121) (203.3,0.0024821) (205.3,0.0023528) (207.3,0.0022247) (209.2,0.0020982) (211.2,0.0019739) (213.2,0.001852) (215.2,0.001733) (217.1,0.0016171) (219.1,0.0015046) (221.1,0.0013958) (223,0.0012908) (225,0.00119) (227,0.0010933) (229,0.0010009)};
\addlegendentry{Xe}
\addplot[orange!90!black, thick] coordinates {(134,0.0011719) (137.8,0.0016083) (141.6,0.0021321) (145.4,0.0027452) (149.2,0.0034475) (153,0.0042361) (156.8,0.0051058) (160.6,0.0060488) (164.4,0.0070555) (168.2,0.0081147) (172,0.0092138) (175.8,0.01034) (179.6,0.011478) (183.4,0.012616) (187.2,0.01374) (190.9,0.014837) (194.7,0.015895) (198.5,0.016906) (202.3,0.017858) (206.1,0.018745) (209.9,0.01956) (213.7,0.020299) (217.5,0.020956) (221.3,0.02153) (225.1,0.022018) (228.9,0.022421) (232.7,0.022741) (236.5,0.022978) (240.3,0.023136) (244.1,0.023218) (247.9,0.023228) (251.7,0.02317) (255.5,0.023048) (259.3,0.022867) (263.1,0.022631) (266.9,0.022345) (270.7,0.022014) (274.5,0.021642) (278.3,0.021234) (282.1,0.020794) (285.9,0.020326) (289.7,0.019834) (293.5,0.019323) (297.3,0.018794) (301.1,0.018252) (304.9,0.0177) (308.7,0.017141) (312.5,0.016576) (316.3,0.016009) (320.1,0.015441) (323.9,0.014875) (327.7,0.014312) (331.5,0.013754) (335.3,0.013203) (339.1,0.012659) (342.9,0.012125) (346.7,0.0116) (350.5,0.011086) (354.3,0.010584) (358.1,0.010094) (361.9,0.0096161) (365.7,0.0091518) (369.5,0.0087012) (373.3,0.0082644) (377.1,0.0078417) (380.9,0.0074331) (384.7,0.0070389) (388.5,0.006659) (392.3,0.0062934) (396.1,0.0059421) (399.9,0.0056048) (403.7,0.0052814) (407.5,0.0049719) (411.3,0.0046758) (415.1,0.0043931) (418.9,0.0041233) (422.7,0.0038663) (426.5,0.0036217) (430.3,0.0033891) (434.1,0.0031683) (437.9,0.0029589) (441.7,0.0027605) (445.5,0.0025727) (449.3,0.0023952) (453.1,0.0022277) (456.9,0.0020696) (460.7,0.0019208) (464.5,0.0017807) (468.3,0.001649) (472.1,0.0015254) (475.9,0.0014095) (479.7,0.0013009) (483.5,0.0011993) (487.3,0.0011044) (491.1,0.0010157)};
\addlegendentry{Ar}
\draw[red, thick] (axis cs:248,1e-3) -- (axis cs:248,1e0);
\end{axis}
\end{tikzpicture}
\caption{Left, recoil spectra in xenon for $m_\chi=500\GeV$, each shown without the LZ acceptance and normalized so that the acceptance-weighted count in the window is one event. The exothermic curves are centered near $160\keV$ and end well below the endothermic curve's endpoint structure; the vertical line at $94.6\keV$ marks the first zero of the $^{131}$Xe form factor, which splits the spectrum into two lobes and places the observed event in the second. Right, argon and xenon at the same $\sigma_p$ for $\delta=-350\keV$; argon peaks at $247\keV$ and exceeds xenon per unit mass, while the endothermic reading forbids argon at any exposure.}
\label{fig:spectra}
\end{figure*}

\begin{table*}[t]
\caption{Exothermic benchmarks. For each $(m_\chi,\delta)$ the cross section is chosen so that Eq.~(\ref{eq:rate}) gives one event in $2.84$ tonne-years over $5.4$--$269.9\keV$, at unit excited fraction. Columns five and six are the ratio of counts in $5.4$--$70\keV$ to those in $70$--$269.9\keV$, and the argon rate and peak at the same $\sigma_p$.}
\label{tab:bench}
\begin{ruledtabular}
\begin{tabular}{cccccccc}
$m_\chi$ & $\delta$ & $\sigma_p$ & Xe peak & $N_{<70}/N_{>70}$ & Ar rate & Ar peak\\
(GeV) & (keV) & (cm$^2$) & (keV) & & (t~yr)$^{-1}$ & (keV)\\
\hline
$300$  & $-300$ & $1.5\times10^{-45}$ & $155$ & $0.106$ & $3.8$ & $206$\\
$500$  & $-275$ & $2.2\times10^{-45}$ & $155$ & $0.188$ & $3.8$ & $195$\\
$500$  & $-350$ & $3.8\times10^{-45}$ & $163$ & $0.011$ & $4.1$ & $247$\\
$800$  & $-350$ & $6.7\times10^{-45}$ & $164$ & $0.009$ & $4.4$ & $250$\\
$1000$ & $-400$ & $1.3\times10^{-44}$ & $172$ & $0.000$ & $4.7$ & $286$\\
\end{tabular}
\end{ruledtabular}
\end{table*}

Table~\ref{tab:bench} collects representative points and Fig.~\ref{fig:spectra} the spectra. Three features are worth separating from the normalization.

\emph{Halo independence.} Varying the escape speed from $500$ to $600\kms$ changes the exothermic count in $70$--$270\keV$ by $0.3\%$, and varying $v_0$ from $220$ to $250\kms$ changes it by $0.5\%$. The endothermic count at $\delta=+340\keV$ changes over the same escape-speed range from zero to $143$ times its value at $544\kms$, and by a factor $2.7$ over the same range of $v_0$. This is the quantitative content of the endpoint tuning, and it is the reason the preferred splitting differs among the published recasts.

\emph{The low-energy window stays empty.} For $\delta=-350\keV$ the counts below $70\keV$ are $1.1\%$ of those above, and for $\delta=-400\keV$ they vanish. The first form-factor lobe, below $94.6\keV$, is populated only through the low-$E_R$ tail of Eq.~(\ref{eq:band}), which requires the fastest particles in the halo. The absence of a low-energy excess in the standard LZ search is therefore predicted rather than accommodated, and it is a genuine constraint on the model, since $|\delta|\lesssim275\keV$ would populate that region.

\emph{The recoil is not at the spectrum peak.} The spectrum peaks near $160\keV$ and the observed event at $248\keV$ lies where the differential rate is smaller by roughly a factor of eight. A single event at that position is unremarkable, in contrast with the endothermic reading, in which the event is required to sit at the endpoint.

\section{Sampling of the halo velocity distribution}
\label{sec:halo}

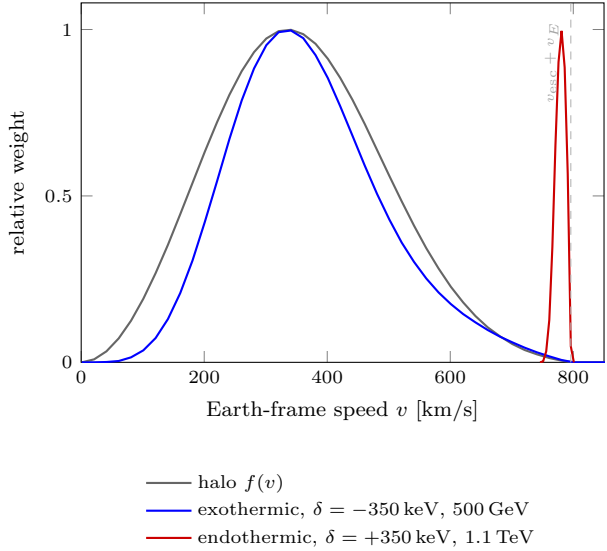
\begin{figure}[t]
\centering
\begin{tikzpicture}
\begin{axis}[
  width=0.8\columnwidth, height=0.55\columnwidth, scale only axis,
  xmin=0, xmax=850, ymin=0, ymax=1.08,
  xlabel={Earth-frame speed $v$ [km/s]}, ylabel={relative weight},
  label style={font=\footnotesize}, tick label style={font=\scriptsize},
  legend style={font=\scriptsize, draw=none, fill=none, at={(0.5,-0.28)}, anchor=north}, legend columns=1,
  legend cell align=left]
\addplot[black!60, thick] coordinates {(1,0.000) (21,0.009) (41,0.033) (61,0.072) (81,0.125) (101,0.191) (121,0.268) (141,0.353) (161,0.445) (181,0.539) (201,0.633) (221,0.722) (241,0.804) (261,0.875) (281,0.932) (301,0.973) (321,0.995) (341,0.999) (361,0.986) (381,0.957) (401,0.912) (421,0.855) (441,0.789) (461,0.716) (481,0.638) (501,0.560) (521,0.484) (541,0.411) (561,0.343) (581,0.282) (601,0.227) (621,0.180) (641,0.139) (661,0.105) (681,0.078) (701,0.055) (721,0.037) (741,0.024) (761,0.013) (781,0.005) (801,0.000) (881,0.000)};
\addlegendentry{halo $f(v)$}
\addplot[blue, thick] coordinates {(1,0.000) (21,0.000) (41,0.001) (61,0.004) (81,0.015) (101,0.036) (121,0.073) (141,0.130) (161,0.208) (181,0.306) (201,0.422) (221,0.545) (241,0.671) (261,0.787) (281,0.883) (301,0.953) (321,0.992) (341,0.997) (361,0.974) (381,0.923) (401,0.854) (421,0.770) (441,0.681) (461,0.591) (481,0.506) (501,0.429) (521,0.360) (541,0.302) (561,0.252) (581,0.210) (601,0.175) (621,0.145) (641,0.120) (661,0.097) (681,0.077) (701,0.060) (721,0.044) (741,0.030) (761,0.018) (781,0.007) (801,0.000) (881,0.000)};
\addlegendentry{exothermic, $\delta=-350\keV$, $500\GeV$}
\addplot[red!80!black, thick] coordinates {(746,0.000) (751,0.003) (756,0.033) (761,0.127) (766,0.346) (771,0.652) (776,0.902) (781,0.996) (786,0.886) (791,0.566) (796,0.048) (801,0.000)};
\addlegendentry{endothermic, $\delta=+350\keV$, $1.1\TeV$}
\draw[gray!60, dashed] (axis cs:796,0) -- (axis cs:796,1.08);
\node[gray!70, font=\tiny, rotate=90, anchor=south east] at (axis cs:790,1.05) {$v_{\rm esc}+v_E$};
\end{axis}
\end{tikzpicture}
\caption{Halo speeds contributing to the LZ-window rate, Eq.~(\ref{eq:w}), each curve scaled to its own peak. The exothermic density is nearly a copy of the halo distribution; the endothermic density is a spike about $20\kms$ wide against the Earth-frame cutoff at $796\kms$.}
\label{fig:sample}
\end{figure}

The halo independence noted in Sec.~\ref{sec:rate} has a simple origin, which is worth displaying because it also fixes the modulation and the argon prediction. The question is which part of the local speed distribution produces the rate in the LZ window. The contributing-speed density is
\begin{equation}
w(v)\propto\frac{f(v)}{v}\sum_i x_i\!\int\! dE_R\,F_i^2\,\varepsilon\,\Theta\!\left[v-v_{\min,i}(E_R)\right],
\label{eq:w}
\end{equation}
with $f(v)$ the Earth-frame speed distribution, $F_i^2$ and $\varepsilon$ the form factor and efficiency at $E_R$, and $v_{\min}$ from Eq.~(\ref{eq:vmin}). Table~\ref{tab:sample} and Fig.~\ref{fig:sample} evaluate it for the halo of Sec.~\ref{sec:rate}, with Earth's orbital motion averaged over the year, giving $v_E\simeq250\kms$ and an Earth-frame maximum of $796\kms$.

\begin{table*}[t]
\caption{Sampling of the halo by the two readings. The quantiles refer to the contributing-speed density of Eq.~(\ref{eq:w}); for the halo distribution itself the median is $349\kms$, with $90\%$ of particles below $537\kms$ and $99\%$ below $679\kms$. $f_{\rm halo}$ is the fraction of halo particles above the lowest contributing speed $v_{\rm low}$. The last two columns give the rate at $v_{\rm esc}=500$ and $600\kms$ relative to $544\kms$; the endothermic entry at $500\GeV$ and $600\kms$ is the quantity quoted as $143$ in Sec.~\ref{sec:rate}, evaluated here with the orbit-averaged $v_E$.}
\label{tab:sample}
\begin{ruledtabular}
\begin{tabular}{lcccccc}
Model & $v_{\rm low}$ (km/s) & $10\%$ / $50\%$ / $90\%$ (km/s) & $f_{\rm halo}$ & Jun/Dec & $R_{500}/R_{544}$ & $R_{600}/R_{544}$\\
\hline
Endothermic, $\delta=+340\keV$, $500\GeV$ & 778 & 785 / 790 / 794 & $1\times10^{-4}$ & $1.5\times10^{4}$ & 0 & 220\\
Endothermic, $\delta=+350\keV$, $1.1\TeV$ & 747 & 768 / 779 / 790 & $1.2\times10^{-3}$ & 18 & 0 & 10\\
Exothermic, $\delta=-350\keV$, $500\GeV$ & 14 & 221 / 357 / 535 & 1.000 & 1.01 & 0.997 & 1.00\\
Exothermic, $\delta=-275\keV$, $500\GeV$ & 1 & 174 / 349 / 590 & 1.000 & 1.02 & 0.989 & 1.01\\
\end{tabular}
\end{ruledtabular}
\end{table*}

The exothermic curve in Fig.~\ref{fig:sample} is nearly a copy of the halo distribution itself, shifted slightly toward higher speeds by the $1/v$ kinematic weighting and the form-factor window, with a median of $357\kms$ against the halo's $349\kms$. The endothermic curve is a spike $20\kms$ wide against the cutoff at $796\kms$. Three consequences follow, and each is a distinct physical statement.

\emph{Which particles are being counted.} In the exothermic model every halo particle can scatter, because $v_{\min}(E_R)$ has a zero at $E_R=(\mu_A/m_A)|\delta|$; the recoil energy is set by the splitting and the halo only broadens the line. In the endothermic model only particles above $v^\star=\sqrt{2\delta/\mu_A}$ scatter at all, and in the LZ window that means the fastest $0.1\%$ ($500\GeV$) to $0.01\%$ ($1.1\TeV$) of the halo. The endothermic rate is therefore a measurement of the population within about $20\kms$ of the escape edge, which is the part of the distribution the Standard Halo Model is least entitled to describe: the truncation there is a modeling convention rather than a measured feature, and Gaia-era determinations place the local escape speed anywhere from $485$ to $580\kms$.

\emph{How the count responds to the halo.} Table~\ref{tab:sample} makes the asymmetry quantitative. Moving $v_{\rm esc}$ from $500$ to $600\kms$ changes the exothermic count by $0.3\%$, and moving $v_0$ from $220$ to $250\kms$ by $0.5\%$; both are smaller than the uncertainty on $\rho_0$. The same $v_{\rm esc}$ excursion takes the endothermic count from zero to $220$ times its reference value at $500\GeV$, and from zero to $10$ times at $1.1\TeV$. Put differently, the exothermic prediction is a statement about $\rho_0\sigma_pf_2$, while the endothermic prediction is a statement about $v_{\rm esc}+v_E$ with the particle physics almost along for the ride. The endothermic reading measures the edge rather than the splitting: with the cross section fixed at weak strength the event selects $\delta\simeq0.93\,\delta_{\max}$, where $\delta_{\max}=\tfrac12\mu_A(v_{\rm esc}+v_E)^2$, so the preferred splitting tracks the assumed escape speed at $0.7\keV$ per $\kms$.

\emph{What this does to the modulation.} Because the endothermic model samples the tip, it sees the full $30\kms$ annual swing of $v_E$ as a swing across the cutoff. The June-to-December ratio is $18$ at $1.1\TeV$ and about $10^4$ at $500\GeV$; the winter is effectively dark. The exothermic model samples the whole distribution, so the swing averages to a percent-level modulation with the usual sign, and at $\delta=-275\keV$ the low-energy lobe reverses it to about $-1.6\%$, as noted in Sec.~\ref{sec:tests}. The strong modulation of the endothermic reading is also found in Ref.~\cite{McCabe}. A second event with a recorded date is therefore close to a decisive discriminator even before any spectral information.

One nuance should be kept in view. The two models sample the halo differently in speed but identically in momentum transfer: both put the LZ event at $q\simeq246\MeV$ in the second lobe of the xenon form factor, so the uncertainty between the Helm and realistic structure functions is common to both and cancels in any ratio between them. The argon prediction of Sec.~\ref{sec:tests} is where that shared uncertainty stops mattering, since argon's first zero lies at $695\keV$ and the event sits in its first lobe.

\section{The mediator is not the \texorpdfstring{$Z$}{Z}}
\label{sec:mediator}

Evaluating Eq.~(\ref{eq:rate}) at weak strength, $\sigma_n^Z=G_F^2\mu_n^2/8\pi=1.86\times10^{-39}\cmsq$ with $Q_W^2$ coherence, gives $8.9\times10^{5}$ events for $m_\chi=500\GeV$ and $\delta=-350\keV$. One event would require an excited fraction $f_2\simeq1\times10^{-6}$, which is not obtained from a thermal history in which the annihilation and downscattering cross sections are both of weak strength; in that case $f_2\simeq\langle\sigma v\rangle_{\rm ann}/\langle\sigma v\rangle_{\rm down}$ is of order unity. A pseudo-Dirac electroweak multiplet is therefore excluded from the exothermic branch, and the required $\sigma_p$ of Table~\ref{tab:bench} points to a dark-sector interaction some six orders of magnitude below weak strength.

For a vector mediator of mass $m_{A'}$ kinetically mixed with hypercharge,
\begin{equation}
\sigma_p=\frac{16\pi\alpha\alpha_D\epsilon^2\mu_p^2}{m_{A'}^4},
\label{eq:sigp}
\end{equation}
so that $\epsilon$ follows once $f_2$ is known. Section~\ref{sec:f2} computes $f_2\ll1$, which raises the required $\sigma_p$ by four to five orders of magnitude relative to Table~\ref{tab:bench} and places $\epsilon$ in the range accessible to dark-photon searches; the numbers are collected in Table~\ref{tab:boltz}.

The momentum transfer also bears on the mediator mass. Since $q$ ranges over $150$--$250\MeV$ across the signal region, a mediator much lighter than $200\MeV$ tilts the spectrum toward its lower edge through the propagator, whereas $m_{A'}\gtrsim q$ leaves the shape of Fig.~\ref{fig:spectra} intact. Spectral shape at a future exposure therefore measures $m_{A'}$ in units of the momentum transfer.

\section{Longevity of the relic excited state}
\label{sec:stability}

The construction requires $\chi_2$ to survive $13.8$~Gyr, which constrains $|\delta|$ from above and the quantum numbers of the pair. The two-body radiative mode $\chi_2\to\chi_1e^+e^-$ is closed for
\begin{equation}
|\delta|<2m_e=1022\keV,
\end{equation}
and $\chi_2\to\chi_1\gamma$ is absent at leading order if the pair is neutral under electromagnetism and carries only the off-diagonal dark current, leaving $\chi_2\to\chi_1\nu\bar\nu$ and $\chi_2\to\chi_1\gamma\gamma$ at rates that are negligible on cosmological time scales. Combining $|\delta|<2m_e$ with $E_R=(\mu_A/m_A)|\delta|=248\keV$ gives $\mu_A>29\GeV$ and hence
\begin{equation}
m_\chi\gtrsim39\GeV,
\end{equation}
with $|\delta|$ running from $1022\keV$ at that mass down to $275\keV$ for $m_\chi\gg m_A$. The mass range is bounded below by a laboratory number rather than chosen.

The same requirement excludes the pseudo-Dirac Higgsino from this branch independently of Sec.~\ref{sec:mediator}. Its excited state decays through the charged-current loop with $\tau\simeq1.87\times10^{-2}\,{\rm s}\,(400\keV/\delta)^3$~\cite{NagataShirai,Visinelli}, so no relic population survives. The exothermic reading thus derives, rather than assumes, that the dark matter is not an electroweak multiplet.

Small $|\delta|$ is technically natural. The limit $\delta\to0$ restores a $U(1)$ acting on the dark fermion, under which the Dirac mass is invariant and the Majorana mass is charged, so radiative corrections to $\delta$ are proportional to $\delta$ itself. This is not so for the Higgsino, where $\delta=m_Z^2(s_W^2/M_1+c_W^2/M_2)$ is a ratio of two large scales and $300\keV$ is obtained by placing the gauginos at $10^7\GeV$~\cite{NagataShirai,Yin}.

\section{Solar capture}
\label{sec:solar}

Pospelov and Ramani have shown that solar capture excludes the endothermic Higgsino interpretation~\cite{PospelovRamani}, and Refs.~\cite{DiMauroShaikh,BoseEtAl} confirm the exclusion with independent treatments of the capture and of the neutrino limits. The Sun accelerates halo particles to $\simeq1400\kms$ in its core and contains heavy elements, so inelastic scattering remains open where terrestrial experiments are kinematically closed; inverting the ten-year IceCube solar limit~\cite{IceCube10yr} they obtain $\delta>566\keV$, three orders of magnitude in annihilation rate away from the LZ-preferred region.

The exothermic branch is not constrained by that argument, but the reason should be stated exactly. Capture is not suppressed; downscattering has no threshold, so $\chi_2$ is captured at least as efficiently as an endothermic state, and the captured population cycles between $\chi_1$ and $\chi_2$ down to the radius at which upscattering closes, since $v^\star\simeq1160\kms$ on iron for $|\delta|=350\keV$ lies below the core escape speed. The evasion rests entirely on the annihilation final state. For a secluded sector~\cite{PospelovRitzVoloshin} the annihilation is $\chi\chi\to A'A'$ with $m_{A'}\ll m_\chi$, and the dark vectors decay to $e^+e^-$, $\mu^+\mu^-$, and pions, all of which stop in the solar interior; the resulting neutrinos have energies of tens of MeV, far below the IceCube threshold. This is the loophole identified in Ref.~\cite{PospelovRamani}, and it is a property of the sector required independently by Secs.~\ref{sec:mediator} and~\ref{sec:stability} rather than an added assumption.

Ref.~\cite{DiMauroShaikh} performs such a calculation for an endothermic pseudo-Dirac benchmark and finds that the two-state kinetics in the Sun suppresses the annihilation rate to five orders of magnitude below the IceCube limit. A dedicated calculation for the exothermic case is nevertheless warranted, both because the captured population is not the halo population and because the loop-induced elastic channels that control thermalization differ from the Higgsino case, being suppressed by $\epsilon^4$ rather than by an electroweak loop.

\section{Tests}
\label{sec:tests}

\emph{Argon.} The sharpest test is a target with a lighter nucleus. The first Helm zero for argon falls at $695\keV$, so at $250\keV$ argon retains $|F|^2=0.081$ against $1.7\times10^{-4}$ for xenon; the loss from $Z^2$ coherence is more than recovered by the form factor and by the larger number of nuclei per unit mass. At the cross section fixed by the LZ event, Table~\ref{tab:bench} gives $3.8$--$4.7$ events per tonne-year in argon, peaking between $195$ and $286\keV$. The endothermic reading predicts nothing in argon at any exposure, since $v^\star=1347\kms$ for $\delta=+350\keV$ and $m_\chi=500\GeV$ exceeds the largest halo speed by a factor of $1.7$. DEAP-3600 and DarkSide-50 have exposures in this range~\cite{DEAP,DarkSide}, and their nuclear-recoil selections were optimized for lower energies, so a dedicated high-recoil reanalysis is the direct test of the scenario. Germanium is the wrong target for the purpose, its first zero falling at $254\keV$, in the middle of the signal region.

\emph{Spectral shape.} The prediction in xenon is a band from roughly $95$ to $270\keV$ peaking near $160\keV$, with a dip at $94.6\keV$ and a smaller first lobe below it whose size grows as $|\delta|$ decreases. A second event should therefore fall below the first rather than at the same energy, which distinguishes this reading from the endothermic one, in which events cluster at the endpoint. The high-energy sideband above the window, which LZ uses for background validation and which Refs.~\cite{RoddSafdiSlatyerXu,DentNewstead} use to constrain both readings, is a further test that a future exposure can apply. The same spectrum continues above the window through the third form-factor lobe. For the benchmarks of Table~\ref{tab:bench} the count expected above $269.9\keV$ is $0.06$ to $0.24$ times the in-window count, taking unit acceptance there, and is consistent with the zero events observed; the fraction grows with the recoil scale $(\mu_A/m_A)|\delta|$, reaching $0.7$ at the heaviest entry of Table~\ref{tab:boltz}. The inelastic-xenon channel, in which the nucleus is left in an excited state, offers a complementary signature~\cite{GuLiTangXu}.

\emph{Annual modulation.} The exothermic amplitude in the $70$--$270\keV$ window is $0.4\%$ at $\delta=-350\keV$, and it reverses sign to $-1.6\%$ at $\delta=-275\keV$, where the low-energy lobe contributes. The endothermic amplitude is $100\%$, with a June-to-December ratio of $5\times10^{3}$; events are confined to roughly May through July. A handful of events would separate the two.

\emph{Other targets.} Lead and tungsten place their first form-factor zeros at $44$ and $53\keV$, so cryogenic calorimeters with those targets probe the second lobe at recoil energies comparable to xenon, with the advantage that exothermic scattering imposes no kinematic requirement on the target mass. Silicon, with its first zero at $1.3\MeV$, retains coherence throughout but suffers from the small recoil $(\mu_A/m_A)|\delta|$.

\section{The excited fraction}
\label{sec:f2}

The excited fraction $f_2$ enters the rate linearly and was set to unity in Table~\ref{tab:bench}. It is not free, and computing it removes the last normalization from the scenario.

\subsection{The dark sector and its rates}

The off-diagonal current $g_DA'_\mu\bar\chi_1\gamma^\mu\chi_2$ permits a vertex only when the two fermion labels differ. Every diagram therefore alternates between the states, which fixes the available channels. Annihilation to two dark vectors proceeds through the opposite state in the propagator, so $\chi_1\chi_1\to A'A'$ and $\chi_2\chi_2\to A'A'$ are allowed and $\chi_1\chi_2\to A'A'$ vanishes; elastic $\chi_1\chi_1\to\chi_1\chi_1$ vanishes at this order, and the only pair conversion is $\chi_2\chi_2\leftrightarrow\chi_1\chi_1$ through $t$- and $u$-channel $A'$ exchange. The absence of a tree-level $\chi_1$ self-interaction removes the halo and cluster constraints that usually accompany a light mediator; the residual effect enters through $\chi_1\chi_1\to\chi_2\chi_2$, which opens above a relative speed $2\sqrt{|\delta|/\mu}$, i.e.\ $709\kms$ at $m_\chi=500\GeV$ and $|\delta|=350\keV$, and is therefore closed in galaxies but open in clusters.

Since the annihilation is secluded, the observed relic density fixes the dark coupling rather than leaving it free,
\begin{equation}
\langle\sigma v\rangle_{\rm ann}=\frac{\pi\alpha_D^2}{m_\chi^2}=2.2\times10^{-26}\,{\rm cm^3\,s^{-1}},
\end{equation}
giving $\alpha_D=0.0122$, $0.0196$, and $0.0245$ at $m_\chi=500$, $800$, and $1000\GeV$. The same secluded-annihilation determination of the dark coupling is used in the endothermic dark-photon analysis of Ref.~\cite{ZhuEtAl}.

For the conversion, single-$A'$ exchange between the pair states is a Yukawa potential of strength $\alpha_D$ appearing off the diagonal of the $\{|11\rangle,|22\rangle\}$ system. In the Born approximation, with relative momenta $k_i$ and $k_f$ related by $k_f^2=k_i^2+2\mu\,\Delta E$ and $\Delta E=2|\delta|$ the energy released,
\begin{equation}
\sigma v_{\rm rel}=\frac{8\pi\mu\alpha_D^2\,k_f}{\left(k_i^2+k_f^2+m_{A'}^2\right)^2-\left(2k_ik_f\right)^2},
\label{eq:svdown}
\end{equation}
the factor of two for identical final particles included. The reverse process follows from the same expression with the momenta exchanged. At threshold $k_f=\sqrt{4\mu|\delta|}=0.59\GeV$ for the central benchmark, so the transfer is fixed once $k_i\to0$. Equation~(\ref{eq:svdown}) is not adequate here, however, and Sec.~\ref{sec:cc} replaces it; the parametric point that survives is that downscattering exceeds annihilation by seven orders of magnitude, which is the origin of the result below.

\subsection{Evolution}

With $Y_i=n_i/s$ and $Y_1+Y_2=Y_{\rm DM}=4.36\times10^{-10}/m_\chi$ fixed after annihilation freeze-out,
\begin{equation}
\frac{dY_2}{d\ln T}=\frac{s(T)}{H(T)}\left[\langle\sigma v\rangle_{\rm down}Y_2^2-\langle\sigma v\rangle_{\rm up}Y_1^2\right],
\label{eq:boltz}
\end{equation}
each average taken over a Maxwellian in $k_i$ at the dark kinetic temperature $T_\chi$. While $T_\chi\gg|\delta|$ the two terms balance and $Y_2/Y_1=e^{-\delta/T_\chi}\simeq1$; once $T_\chi$ falls below $|\delta|$ the upscattering shuts off and the depletion is one-way until it freezes. Because the coefficient $s/(HT)$ is nearly constant, the integral is controlled by its upper limit and
\begin{equation}
f_2\simeq\left[\langle\sigma v\rangle_{\rm down}\,\frac{s}{HT}\,T_\star\,Y_{\rm DM}\right]^{-1},
\label{eq:f2est}
\end{equation}
with $T_\star\simeq\sqrt{|\delta|\,T_{\rm kd}}$ the photon temperature at which $T_\chi$ reaches $|\delta|$. The dark states remain kinetically coupled through the thermal $A'$ population until it disappears, so $T_{\rm kd}\simeq m_{A'}/15$.

Integrating Eq.~(\ref{eq:boltz}) with the Born rate gives $f_2\simeq10^{-4}$, with Eq.~(\ref{eq:f2est}) reproducing it to a factor of three. The dependence on the least certain input is mild: varying $T_{\rm kd}$ from $5$ to $100\MeV$ changes $f_2$ by a factor of five, and varying $|\delta|$ from $275$ to $1000\keV$ changes it by two. Since $f_2\propto\alpha_D^{-2}$ while $\sigma_p\propto\alpha_D\epsilon^2$, the inferred mixing scales only as $\epsilon\propto\alpha_D^{1/2}$, so an $O(1)$ error in the relic normalization of $\alpha_D$ is not important. The Born rate itself is not adequate, and the next subsection replaces it.

\subsection{Coupled-channel conversion}
\label{sec:cc}

At the couplings fixed by the relic abundance, $\alpha_Dm_\chi/k_f\simeq10$, so the ladder of $A'$ exchanges is unsuppressed and Eq.~(\ref{eq:svdown}) cannot be trusted. The structure of the correction is visible before any computation. Rotating to $|\pm\rangle=(|11\rangle\pm|22\rangle)/\sqrt2$ diagonalizes the off-diagonal potential into an attractive and a repulsive Yukawa of equal strength, and in the limit $|\delta|\to0$ the conversion becomes a phase-shift difference,
\begin{equation}
\sigma_0=\frac{\pi}{k^2}\sin^2\!\left(\delta_+-\delta_-\right),
\label{eq:eigen}
\end{equation}
bounded by $\pi/k^2$ for every value of the coupling. The Born expression is the small-phase limit of Eq.~(\ref{eq:eigen}) with a perturbative phase, and it carries no such bound; at $m_{A'}=0.6\GeV$ and $k=0.05\GeV$ it exceeds the $s$-wave unitarity limit by a factor of nine.

We therefore solve the radial system
\begin{align}
u_1''+\left[k_1^2-\tfrac{l(l+1)}{r^2}\right]u_1&=2\mu V(r)\,u_2,\nonumber\\
u_2''+\left[k_2^2-\tfrac{l(l+1)}{r^2}\right]u_2&=2\mu V(r)\,u_1,
\label{eq:radial}
\end{align}
with $V(r)=-\alpha_De^{-m_{A'}r}/r$ and $k_1^2=k_2^2+4\mu|\delta|$, matching the regular solutions at large $r$ to obtain the reaction matrix $K$, the scattering matrix $S=(1+iK)(1-iK)^{-1}$, and
\begin{equation}
\sigma_{2\to1}=\frac{\pi}{k_2^2}\sum_l(2l+1)\left|S_{12}\right|^2,
\end{equation}
summed to $l=16$. Four checks control the numerics. The reaction matrix is symmetric to one part in $10^{9}$ and $S$ is unitary to one part in $10^{10}$; taking $\alpha_D\to10^{-4}$ recovers the analytic Born partial waves to four digits; the Born partial-wave sum reproduces the closed form of Eq.~(\ref{eq:svdown}) exactly; and the suppressed cases are converged to seven digits against the inner and outer radii and the integration tolerance.

The measured phase difference is $0.03$ to $1.3$ rad across the relevant momenta, not large, so Eq.~(\ref{eq:eigen}) is far below the Born estimate. Thermally averaged, the conversion rate falls by roughly an order of magnitude, from $3.6\times10^{-18}$ to $3.6\times10^{-19}\,{\rm cm^3\,s^{-1}}$ at $T_\chi=|\delta|$ for $m_{A'}=1\GeV$. Downscattering still exceeds annihilation by $1.4\times10^{7}$, so the depletion picture stands and only its magnitude moves.

\subsection{Consequences}

\begin{table*}[t]
\caption{Results of Eqs.~(\ref{eq:boltz}) and (\ref{eq:radial}). $\alpha_D$ is fixed by the relic abundance and $T_{\rm kd}=m_{A'}/15$. The Born column is shown for comparison. The cross section normalizes the sum of the two components to one LZ event, and $\epsilon$ follows from Eq.~(\ref{eq:sigp}). The last column is the predicted argon rate; the exothermic share is unity at every entry, the endothermic channel being kinematically closed.}
\label{tab:boltz}
\begin{ruledtabular}
\begin{tabular}{cccccccc}
$m_\chi$ & $|\delta|$ & $m_{A'}$ & $\alpha_D$ & $f_2$ (Born) & $f_2$ (coupled) & $\sigma_p$ (cm$^2$) & $\epsilon$\\
(GeV) & (keV) & (GeV) & & & & & \\
\hline
$500$  & $350$ & $0.6$ & $0.0122$ & $9.5\times10^{-5}$ & $2.2\times10^{-2}$ & $1.7\times10^{-43}$ & $1.2\times10^{-7}$\\
$500$  & $350$ & $0.8$ & $0.0122$ & $1.5\times10^{-4}$ & $2.4\times10^{-3}$ & $1.6\times10^{-42}$ & $6.5\times10^{-7}$\\
$500$  & $350$ & $1.0$ & $0.0122$ & $2.2\times10^{-4}$ & $2.1\times10^{-3}$ & $1.8\times10^{-42}$ & $1.1\times10^{-6}$\\
$500$  & $350$ & $2.0$ & $0.0122$ & $1.2\times10^{-3}$ & $3.8\times10^{-2}$ & $1.0\times10^{-43}$ & $1.0\times10^{-6}$\\
$500$  & $350$ & $3.0$ & $0.0122$ & $3.9\times10^{-3}$ & $2.7\times10^{-3}$ & $1.4\times10^{-42}$ & $8.6\times10^{-6}$\\
$800$  & $450$ & $1.0$ & $0.0196$ & $8.4\times10^{-5}$ & $4.7\times10^{-3}$ & $3.5\times10^{-42}$ & $1.2\times10^{-6}$\\
$1000$ & $500$ & $1.0$ & $0.0245$ & $5.6\times10^{-5}$ & $8.0\times10^{-3}$ & $4.9\times10^{-42}$ & $1.3\times10^{-6}$\\
$1000$ & $500$ & $2.0$ & $0.0245$ & $2.1\times10^{-4}$ & $1.4\times10^{-2}$ & $2.9\times10^{-42}$ & $3.9\times10^{-6}$\\
\end{tabular}
\end{ruledtabular}
\end{table*}

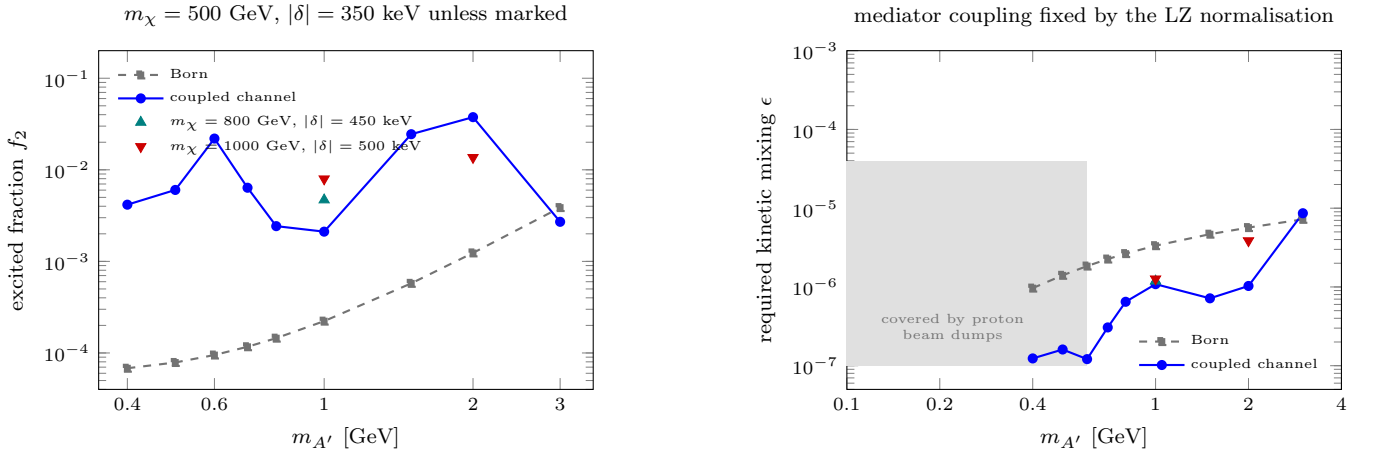
\begin{figure*}[t]
\centering
\begin{tikzpicture}
\begin{axis}[
  width=0.365\textwidth, height=0.25\textwidth, scale only axis,
  xmode=log, ymode=log, xmin=0.35, xmax=3.5, ymin=4e-5, ymax=2e-1,
  xtick={0.4,0.6,1,2,3}, xticklabels={0.4,0.6,1,2,3},
  xlabel={$m_{A'}$ [GeV]}, ylabel={excited fraction $f_2$},
  title={\footnotesize $m_\chi=500$ GeV, $|\delta|=350$ keV unless marked},
  label style={font=\footnotesize}, tick label style={font=\scriptsize},
  legend style={font=\tiny, draw=none, fill=none, at={(0.02,0.98)}, anchor=north west},
  legend cell align=left]
\addplot[black!55, dashed, thick, mark=square*, mark size=1.3] coordinates {(0.4,6.811e-05) (0.5,7.872e-05) (0.6,9.494e-05) (0.7,0.0001169) (0.8,0.0001451) (1,0.0002226) (1.5,0.0005756) (2,0.001237) (3,0.003869)};
\addlegendentry{Born}
\addplot[blue, thick, mark=*, mark size=1.5] coordinates {(0.4,0.004161) (0.5,0.006028) (0.6,0.02194) (0.7,0.006373) (0.8,0.002426) (1,0.002112) (1.5,0.02448) (2,0.03759) (3,0.002705)};
\addlegendentry{coupled channel}
\addplot[teal, only marks, mark=triangle*, mark size=2.2] coordinates {(1,0.004691)};
\addlegendentry{$m_\chi=800$ GeV, $|\delta|=450$ keV}
\addplot[red!80!black, only marks, mark=triangle*, mark options={rotate=180}, mark size=2.2] coordinates {(1,0.007958) (2,0.01364)};
\addlegendentry{$m_\chi=1000$ GeV, $|\delta|=500$ keV}
\end{axis}
\end{tikzpicture}%
\hfill%
\begin{tikzpicture}
\begin{axis}[
  width=0.365\textwidth, height=0.25\textwidth, scale only axis,
  xmode=log, ymode=log, xmin=0.1, xmax=4, ymin=5e-8, ymax=1e-3,
  xtick={0.1,0.2,0.4,1,2,4}, xticklabels={0.1,0.2,0.4,1,2,4},
  xlabel={$m_{A'}$ [GeV]}, ylabel={required kinetic mixing $\epsilon$},
  title={\footnotesize mediator coupling fixed by the LZ normalisation},
  label style={font=\footnotesize}, tick label style={font=\scriptsize},
  legend style={font=\tiny, draw=none, fill=none, at={(0.98,0.02)}, anchor=south east},
  legend cell align=left]
\fill[black!12] (axis cs:0.1,1e-7) rectangle (axis cs:0.6,4e-5);
\node[black!55, font=\tiny, align=center] at (axis cs:0.22,3e-7) {covered by proton\\beam dumps};
\addplot[black!55, dashed, thick, mark=square*, mark size=1.3] coordinates {(0.4,9.668e-07) (0.5,1.405e-06) (0.6,1.842e-06) (0.7,2.26e-06) (0.8,2.65e-06) (1,3.342e-06) (1.5,4.677e-06) (2,5.672e-06) (3,7.216e-06)};
\addlegendentry{Born}
\addplot[blue, thick, mark=*, mark size=1.5] coordinates {(0.4,1.237e-07) (0.5,1.606e-07) (0.6,1.212e-07) (0.7,3.061e-07) (0.8,6.48e-07) (1,1.085e-06) (1.5,7.171e-07) (2,1.029e-06) (3,8.63e-06)};
\addlegendentry{coupled channel}
\addplot[teal, only marks, mark=triangle*, mark size=2.2] coordinates {(1,1.187e-06)};
\addplot[red!80!black, only marks, mark=triangle*, mark options={rotate=180}, mark size=2.2] coordinates {(1,1.264e-06) (2,3.861e-06)};
\end{axis}
\end{tikzpicture}
\caption{Left, the relic excited fraction from Eq.~(\ref{eq:boltz}) with Born and with coupled-channel conversion rates. The oscillation in $m_{A'}$ is the phase difference of Eq.~(\ref{eq:eigen}) entering a cosmological quantity. The filled triangles are the coupled-channel results for the two heavier benchmarks of Table~\ref{tab:boltz}, pointing up for $m_\chi=800\GeV$ at $|\delta|=450\keV$ and down for $1000\GeV$ at $500\keV$; the curves are $500\GeV$ at $350\keV$ throughout, and the triangles carry the same meaning in both panels, where the right one is not relabeled. Right, the kinetic mixing required to reproduce one LZ event, from Eq.~(\ref{eq:sigp}) at the same $f_2$; the shaded rectangle marks the region covered by proton beam dumps, which pushes the viable mediator above $0.6\GeV$.}
\label{fig:f2}
\end{figure*}

Table~\ref{tab:boltz} and Fig.~\ref{fig:f2} collect the outcome. Three consequences follow.

\emph{The mediator sits in a searched region.} With $f_2=2\times10^{-3}$ to $4\times10^{-2}$ the required $\sigma_p$ is $10^{-43}$--$10^{-42}\cmsq$ and the mixing is $\epsilon\simeq1\times10^{-6}$ at $m_{A'}=1\GeV$, rising to $9\times10^{-6}$ at $3\GeV$. Proton beam dumps cover roughly $10^{-7}\lesssim\epsilon\lesssim4\times10^{-5}$ for $m_{A'}$ below about $0.6\GeV$~\cite{NuCal,NA62dump}, which excludes the lighter benchmarks and leaves $m_{A'}\gtrsim0.6\GeV$, where existing searches require only $\epsilon\lesssim10^{-3}$~\cite{DarkPhotonReview}. The coupled-channel treatment moves $\epsilon$ down by a factor of three relative to Born and therefore closer to that boundary, so an overlay against the current compilation is needed before it is treated as sharp.

\emph{The ground state is not spectator.} With $f_2\lesssim4\times10^{-2}$ the remaining $98\%$ or more of the halo is $\chi_1$, which upscatters endothermically through the same operator and at the same $\sigma_p$. The two components must therefore be normalized together, and the endothermic contribution is not negligible at the larger $\sigma_p$ now required. At $m_\chi=500\GeV$ and $|\delta|=300\keV$ the endothermic channel supplies $86\%$ of the count, at $|\delta|=350\keV$ it is closed because $v^\star=802\kms$ exceeds the halo cutoff, and the entries of Table~\ref{tab:boltz} are pure exothermic. The requirement that the endothermic component not overshoot excludes $|\delta|\lesssim300\keV$ at $m_\chi=500\GeV$, reinforcing the bound already obtained in Sec.~\ref{sec:rate} from the low-energy window. The published endothermic interpretations are recovered as the small-splitting limit of the same model, with the two readings separated by whether $v^\star$ lies inside the halo.

\emph{The argon prediction survives.} The exothermic rate is normalized to the LZ event, so the argon prediction of Sec.~\ref{sec:tests} is unchanged in the pure-exothermic entries, $4.1$ to $6.3$ events per tonne-year. Where the endothermic component contributes, the argon rate falls in proportion to the exothermic share, so an argon measurement determines that share directly.

Attenuation is negligible. For mean terrestrial composition the nuclear cross section is $3\times10^{-37}\cmsq$ at the central benchmark and the probability that a $\chi_2$ downscatters while crossing the Earth is below $10^{-4}$.

Two limitations remain. The identical-particle and spin factors in Eq.~(\ref{eq:radial}) are treated as in the Born expression, an $O(1)$ normalization common to both that does not affect the comparison. More importantly, $f_2$ oscillates with $m_{A'}$ by an order of magnitude through the phase difference of Eq.~(\ref{eq:eigen}), so the prediction at any single mediator mass carries that spread; only the range $f_2=10^{-3}$ to $10^{-1}$, and the resulting $\epsilon\simeq10^{-7}$ to $10^{-5}$, should be regarded as the output.

\section{Summary}

Downscattering reproduces the LZ event without placing the splitting at the xenon kinematic ceiling, and the resulting count is stable to $0.3\%$ against the halo escape speed where the endothermic count varies by two orders of magnitude. The reading is internally constrained rather than free. A weak-strength coupling overproduces the event by $8.9\times10^5$, so the mediator is a dark-sector one; longevity of the relic excited state requires $|\delta|<2m_e$ and hence $m_\chi\gtrsim39\GeV$, and it excludes the Higgsino, whose excited state lives $3\times10^{-2}$ s; and the absence of a low-energy excess follows from the kinematics for $|\delta|\gtrsim300\keV$. Solving the coupled Boltzmann system removes the remaining normalization. The relic abundance fixes $\alpha_D$ and downscattering exceeds annihilation by seven orders of magnitude, so the excited state is depleted. The conversion has to be treated nonperturbatively, since the off-diagonal potential makes the rate a phase-shift difference bounded by unitarity while the Born expression is not; the coupled-channel solution lies an order of magnitude below Born and gives $f_2=2\times10^{-3}$ to $4\times10^{-2}$, oscillating with $m_{A'}$ through that phase difference. One event then requires $\sigma_p\simeq2\times10^{-42}\cmsq$ and $\epsilon\simeq1\times10^{-6}$ at $m_{A'}=1\GeV$, in a searched region, with beam-dump coverage below $0.6\GeV$ excluding the lighter benchmarks.

A consequence of $f_2\ll1$ is that the ground state carries the halo and upscatters endothermically through the same operator, so the two components are normalized together. That requirement excludes $|\delta|\lesssim300\keV$ and identifies the published endothermic interpretations as the small-splitting limit of this model, the two readings separated by whether $v^\star$ lies inside the halo.

The scenario is falsifiable with existing data. Argon should show $4$ to $6$ events per tonne-year near $250\keV$, where the endothermic reading is kinematically forbidden, and the annual modulation should be at the percent level rather than complete. What would sharpen the cosmology further is a treatment of the conversion that keeps the spin structure and the resonance region of Eq.~(\ref{eq:eigen}), since the order-of-magnitude oscillation of $f_2$ with $m_{A'}$ is the dominant remaining spread.

\begin{acknowledgments}
V.B. gratefully acknowledges support from the William F. Vilas Estate. H.B. gratefully acknowledges support from the Avenir Foundation.
\end{acknowledgments}

\section*{Addendum: relation to other interpretations}

The LZ event has drawn many interpretations in the days since the preprint appeared, and this Addendum places the present reading among them. Most are endothermic~\cite{Nomura,DiMauroEdge,Yamashita,ChattopadhyayDasPuriRoy,SmirnovGriffithBeacom,DuWang,McCabe,WangXiao,YangWuTsaiFan,KotlarskiKowalskaSessolo,DasKarmakarMahapatraPaul,OkadaSeto,AhmedLeontaris,DuHuangXie,BandyopadhyayBorahBorah,BorahSahooSahuSharma,BisalCaoLi,CheungKangKumar,YuanZhangCaoFengYang,ZhuEtAl,LeeRandall,LeeYoun,HMLee,AsadiBatzFoxHomillerKribs}, with a nearly pure Higgsino, a vectorlike electroweak doublet, or a pseudo-Dirac state of a dark sector upscattering to a partner state heavier by a few hundred keV. All of them share the first two conditions stated in Sec.~\ref{sec:intro}, and three tests of that reading have been noted. The empty high-energy sideband above the LZ window disfavors a thermal Higgsino~\cite{RoddSafdiSlatyerXu,DentNewstead}, the annual modulation must be large~\cite{McCabe}, and solar capture excludes the thermal Higgsino~\cite{PospelovRamani,DiMauroShaikh,BoseEtAl}. The exothermic reading meets each of these differently. Its xenon spectrum lies mainly below $270\keV$ (Sec.~\ref{sec:tests}), its modulation is at the percent level (Sec.~\ref{sec:halo}), and its solar neutrino signal is suppressed by the secluded annihilation (Sec.~\ref{sec:solar}).

Two papers apply exothermic dark matter to the event. Ref.~\cite{deLima} realizes it in an inelastic dark-photon model and finds a preferred region $m_\chi\sim30$--$200\GeV$ with $|\delta|\sim0.5$--$1\MeV$, and Ref.~\cite{DentNewstead} fits the event and finds the mass unconstrained by the fit alone, since a heavy candidate can hide its peak above the region of interest; including the empty high-energy sideband that LZ uses for background validation removes that freedom, because a broadly peaked spectrum overshoots the zero events observed there. An exothermic reading is thereby pushed toward the sharper peaks produced by lighter dark matter with larger splittings. The present analysis overlaps with both in the mechanism and differs in the mass range it selects, in the coupled-channel computation of the relic excited fraction that fixes the normalization, and in the bound on the splitting from upscattering of the ground state.

Elastic readings through spin-dependent or momentum-dependent operators~\cite{DiMauroEdge,Unwin,ElahiSchwaller} have also been proposed, as have readings that do not involve halo dark matter at all, among them boosted dark matter~\cite{LiangLiuTranXu,AlhazmiEtAl,KannikeRaidalStrumia}, absorption of a fermionic dark matter particle~\cite{LouLu}, atmospheric-neutrino upscattering~\cite{JeesunMajumdar}, and neutron disappearance~\cite{AghaieStrumia}. These are outside the scope of this paper.

\end{document}